\documentclass[aps,prl,superscriptaddress,twocolumn,nofootinbib]{revtex4} 
\usepackage{graphicx}

\begin{document}
\title{$\mathbf{D}$--$\mathbf{\bar D}$ 
  MIXING AND NEW PHYSICS:\\ 
GENERAL CONSIDERATIONS AND CONSTRAINTS ON THE MSSM}

\author{M.~Ciuchini}
\affiliation{Dipartimento di Fisica, Universit\`a di Roma Tre 
and INFN, Sezione di Roma Tre, Via della Vasca Navale 84, I-00146
Roma, Italy}
\author{E.~Franco}
\affiliation{Dipartimento di Fisica, Universit\`a di Roma ``La
  Sapienza''  and INFN, 
  Sezione di Roma, P.le A. Moro 2, I-00185 Rome, Italy}
\author{D.~Guadagnoli}
\affiliation{Physik-Department T31, TU-M{\"u}nchen, D-85748 Garching,
  Germany} 
\author{V.~Lubicz}
\affiliation{Dipartimento di Fisica, Universit\`a di Roma Tre 
and INFN, Sezione di Roma Tre, Via della Vasca Navale 84, I-00146
Roma, Italy}
\author{M.~Pierini}
\affiliation{Department of Physics, University of Wisconsin, Madison,
  WI 53706, USA}
\author{V.~Porretti}
\affiliation{Departament de F\'{\i}sica Te\`orica and IFIC, Universitat de
Val\`encia-CSIC, E-46100, Burjassot, Spain}
\author{L.~Silvestrini}
\affiliation{Dipartimento di Fisica, Universit\`a di Roma ``La
  Sapienza''  and INFN, 
  Sezione di Roma, P.le A. Moro 2, I-00185 Rome, Italy}

\begin{abstract}
  Combining the recent experimental evidence of $D$--$\bar
  D$ mixing, we extract model-independent information on the mixing
  amplitude and on its CP-violating phase. Using this information, we
  present new constraints on the flavour structure of up-type squark
  mass matrices in supersymmetric extensions of the Standard Model.
\end{abstract}

\maketitle

The study of meson oscillations represents one of the most powerful probes
of New Physics (NP) currently available. The $K$ and $B_d$ systems are
very well studied experimentally and all the measurements performed up
to now are compatible with the Standard Model (SM) expectation,
although there is still room for NP which could be revealed with
improved theoretical tools and experimental facilities hopefully
available in the future~\cite{UTnp06,CDR}.

As far as the $B_s$ is concerned, the experimental evidence of
oscillation was found only recently at the TeVatron~\cite{DMS}. While
the oscillation frequency is already very well known, information on
the phase of the mixing amplitude is still quite vague, leaving ample
room for experimental improvements expected from hadronic colliders.

All this experimental information allows to put model-independent
constraints on NP contributions to the mixing amplitudes involving
down-type quarks~\cite{UTnp06}.  These constraints already induce highly
non-trivial bounds on the flavour structure of many extensions of the
SM. In particular, considering the Minimal Supersymmetric Standard
Model (MSSM), the flavour properties of the down-type squark mass
matrices have already been thoroughly analyzed~\cite{susydf2}. 

On the other hand, up to now no evidence was found of oscillations of
mesons involving up-type quarks. Correspondingly, the off-diagonal
entries of up-type squark mass matrices were only weakly
bounded~\cite{ddbarsusy,nirraz}. It is remarkable that one of the proposed
mechanisms to explain the flavour structure of the MSSM and to
suppress the unwanted SUSY contributions to Flavour-Changing Neutral
Current (FCNC) processes, namely alignment of quark and squark mass
matrices~\cite{alignment}, naturally produces sizable effects in the
up-type sector. In the absence of stringent experimental information,
these models were not tightly constrained~\cite{nirraz}.

Very recently, BaBar~\cite{babarDD} and Belle~\cite{bellexy,belleycp}
independently reported evidence for $D$--$\bar D$ mixing. In this
letter we use this information, combined with previous constraints on
D mixing~\cite{otherxy,otherycp,cleo,RMall}, to put model-independent bounds on the
mixing amplitude and to constrain the relevant entries of the up-type
squark mass matrices. To fulfil this task we use the mass-insertion
approximation. Treating off-diagonal sfermion mass terms as
interactions, we perform a perturbative expansion of FCNC amplitudes
in terms of mass insertions.  The lowest non-vanishing order of this
expansion gives an excellent approximation to the full result, given
the tight experimental constraints on flavour changing mass
insertions. It is most convenient to work in the super-CKM basis, in
which all gauge interactions carry the same flavour dependence as SM
ones. In this basis, we define the mass insertions
$\left(\delta^u_{12}\right)_{AB}$ as the off-diagonal mass terms
connecting up-type squarks of flavour $u$ and $c$ and helicity $A$ and
$B$, divided by the average squark mass.

Let us first discuss the recent experimental novelties. BaBar studied
$D^0 \to K^+ \pi^-$ and $\bar D^0 \to K^- \pi^+$ decays as a function
of the proper time of the $D$ mesons. Assuming no CP violation in
mixing, which is safe in the SM, this analysis allows
to measure the parameters $x^{\prime 2}$ and $y^\prime$, defined in terms
of the mixing parameters $x$ and $y$ through the relations:
$$  x^\prime = ~x \cos \delta_{K\pi} + y \sin \delta_{K\pi},\quad
  y^\prime = -x \sin \delta_{K\pi} + y \cos \delta_{K\pi} ,
$$
where $\delta_{K\pi}$ is the relative strong phase between the
Cabibbo-suppressed $D^0 \to K^+ \pi^-$ decays and the Cabibbo-favoured
$D^0 \to K^- \pi^+$ ones. This phase has been recently measured by
CLEO-c~\cite{cleo}. From a fit to $D^0$ and $\bar D^0$ decays, BaBar
is able to exclude the point $x^{\prime 2} = y^\prime = 0$ (which
corresponds to the no-mixing scenario) with a $3.9\sigma$ significance
(including systematic effects). In addition, the BaBar collaboration
fitted separately the parameters $x^{\prime 2}_\pm$ and $y^\prime_\pm$
of $D^0 \to K^\pm \pi^\mp$ decays allowing for CP violation.

Belle directly determines $x$ and $y$, studying the $D^0 \to K^0_S
\pi^+ \pi^-$ Dalitz plot. In this way, one can separately measure the
mixing parameters and the strong phase. Even though this analysis is
not precise enough to claim the observation of $D$--$\bar D$ mixing,
it allows to disentangle mixing parameters from the strong phase
$\delta_{K\pi}$, when $D^0 \to K^0_S \pi^+ \pi^-$ and $D^0 \to K \pi$
results are combined.

Belle also found evidence of $D$--$\bar D$ mixing, observing a
deviation from zero (at $3.2\sigma$ including the systematic error) of
$y_\mathrm{CP} = \frac{\tau(D^0 \to K^- \pi^+)}{\tau(D^0 \to
  f_\mathrm{CP})}-1$ and in addition measured the CP asymmetry
$A_\Gamma=(\Gamma(D\to KK)-\Gamma(\bar D\to KK))/(\Gamma(D\to
KK)+\Gamma(\bar D\to KK))$.

We assume that CP violation can occur in mixing but not in decay
amplitudes, since the latter are dominated by SM tree-level
contributions. Therefore, we assume that $\Gamma_{12}$ is real. Our
aim is to determine the parameters $\vert M_{12}\vert e^{-i \Phi_{12}}$ and
  $\Gamma_{12}$ from the available experimental data. One can write
  the following relations~\cite{Raz}:
\begin{eqnarray}
  && \vert M_{12} \vert = \frac{1}{\tau_D } \sqrt{\frac{x^2+\delta^2
      y^2}{4(1-\delta^2)}}\,,\quad
  \vert \Gamma_{12} \vert= \frac{1}{\tau_D }\sqrt{\frac{y^2+\delta^2
      x^2}{1-\delta^2}}\,, \nonumber \\
  && \sin \Phi_{12} = \frac{\vert \Gamma_{12}\vert^2 + 4 \vert
    M_{12}\vert^2 - (x^2+y^2)\vert q/p\vert^2/\tau_D^2}{4 \vert M_{12}
    \Gamma_{12}\vert}\,, \nonumber \\
  &&\phi = \arg (y+i \delta x)\,,\quad
  y^\prime_\pm   =
  \left\vert
    \frac{q}{p}
  \right\vert^{\pm 1}(y^\prime\cos \phi \mp x^\prime \sin
  \phi)\,, \nonumber \\ 
  && x^{\prime 2}_\pm = \left\vert
    \frac{q}{p}
  \right\vert^{\pm 2}(x^\prime\cos \phi \pm y^\prime \sin
  \phi)^2\,,\quad R_M =\frac{x^2+y^2}{2}\,,\nonumber \\ 
  && y_\mathrm{CP} =
  \left(
    \left\vert
      \frac{q}{p}
    \right\vert + \left\vert
      \frac{p}{q}
    \right\vert
  \right) \frac{y}{2} \cos \phi- \left(
    \left\vert
      \frac{q}{p}
    \right\vert - \left\vert
      \frac{p}{q}
    \right\vert
  \right) \frac{x}{2}\sin \phi\,, \nonumber \\
  && A_\Gamma =  \left(
    \left\vert
      \frac{q}{p}
    \right\vert - \left\vert
      \frac{p}{q}
    \right\vert
  \right) \frac{y}{2} \cos \phi- \left(
    \left\vert
      \frac{q}{p}
    \right\vert + \left\vert
      \frac{p}{q}
    \right\vert
  \right) \frac{x}{2}\sin \phi\,, \nonumber
\end{eqnarray}
where $\delta=\vert p \vert^2 - \vert q \vert^2$ and $\phi$ is the phase of the mixing parameter $q/p$. We fit for $\vert
M_{12}\vert$, $\vert \Gamma_{12}\vert$ and $\Phi_{12}$ using the
experimental inputs listed in Table~\ref{tab:exp}, taking into account
the correlations between $y^\prime_\pm$ and $x^{\prime 2}_\pm$ in the
BaBar results. Notice that all observables can be written in terms of
$\vert M_{12}\vert$, $\vert \Gamma_{12}\vert$ and $\Phi_{12}$. 

\begin{table}[!tb]
\begin{center}
\begin{tabular}{lcr}
\hline
Parameter & Value & Ref. \\
\hline
$x^{\prime 2}_+$ & $(-0.24 \pm 0.43 \pm 0.30)\cdot 10^{-3}$ &
\cite{babarDD} \\
$x^{\prime 2}_-$ & $(-0.20 \pm 0.41 \pm 0.29)\cdot 10^{-3}$ &
\cite{babarDD} \\
$y^\prime_+$ & $(9.8 \pm 6.4 \pm 4.5)\cdot 10^{-3}$ & \cite{babarDD} \\
$y^\prime_-$ & $(9.6 \pm 6.1 \pm 4.3)\cdot 10^{-3}$ & \cite{babarDD} \\
$x$ & $(7.9 \pm 3.4)\cdot 10^{-3}$ & \cite{bellexy,otherxy} \\
$y$ & $(3.4 \pm 2.8)\cdot 10^{-3}$ & \cite{bellexy,otherxy} \\
$\phi$ ($^\circ$) & $-12 \pm 18$ & \cite{bellexy,otherxy} \\
$\vert q/p \vert$ & $0.88 \pm 0.31$ & \cite{bellexy}\\
$y_\mathrm{CP}$ & $(11.2 \pm 3.2)\cdot 10^{-3}$ & \cite{belleycp,otherycp} \\
$A_\Gamma$ & $(-1.7 \pm 3.0)\cdot 10^{-3}$ & \cite{belleycp,otherycp} \\
$\cos\delta_{K\pi}$ & $1.09\pm 0.66$ & \cite{cleo} \\
$R_M$ & $(21\pm 11)\cdot 10^{-3}$ & \cite{RMall} \\
$\tau_D$ (ps) & $0.4101 \pm 0.0015$ & \cite{pdg} \\
\hline
\end{tabular}
\end{center}
\caption{Experimental results used in our analysis. For $A_\Gamma$,
  $y_\mathrm{CP}$ and $R_M$ we have used the Heavy Flavor Averaging
  Group (HFAG) averages as of May 2007. For $x$, $y$ and $\phi$ we
  have performed our own combination of experimental results as the
  HFAG averages are obtained assuming no CP violation.}
\label{tab:exp}
\end{table}

\begin{figure}[htb!]
\begin{center}
\includegraphics[width=0.23\textwidth]{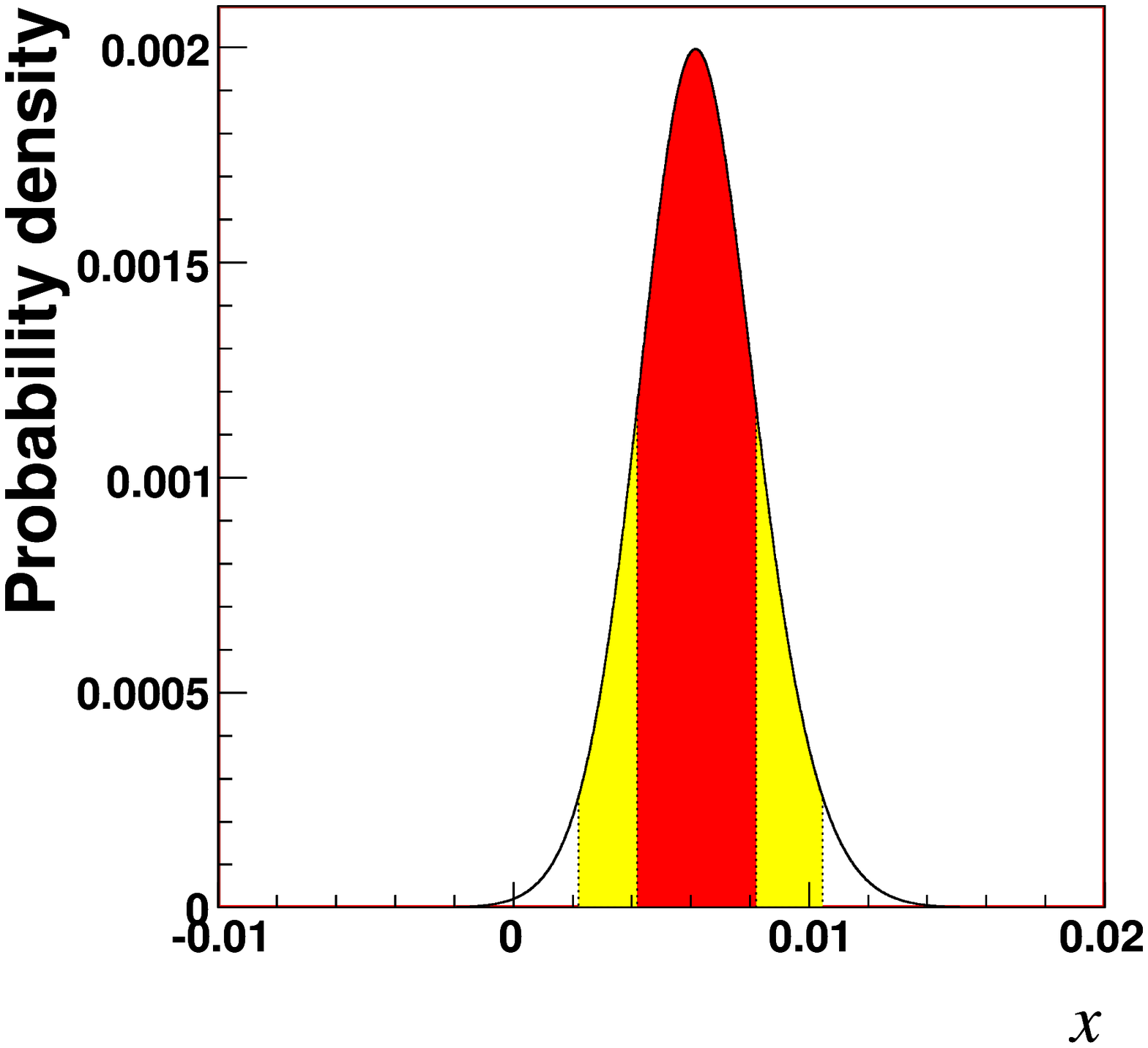}
\includegraphics[width=0.23\textwidth]{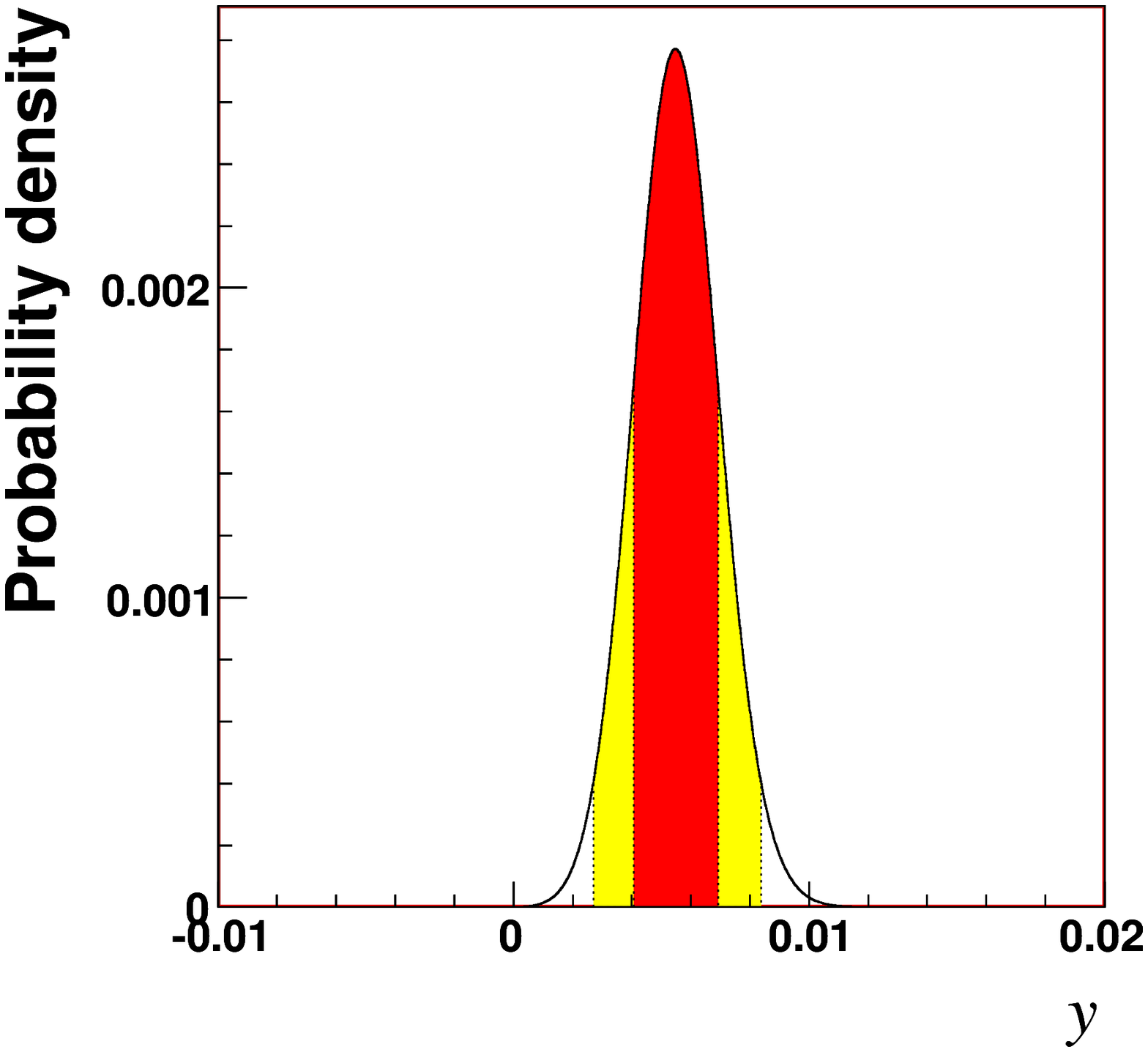}
\includegraphics[width=0.23\textwidth]{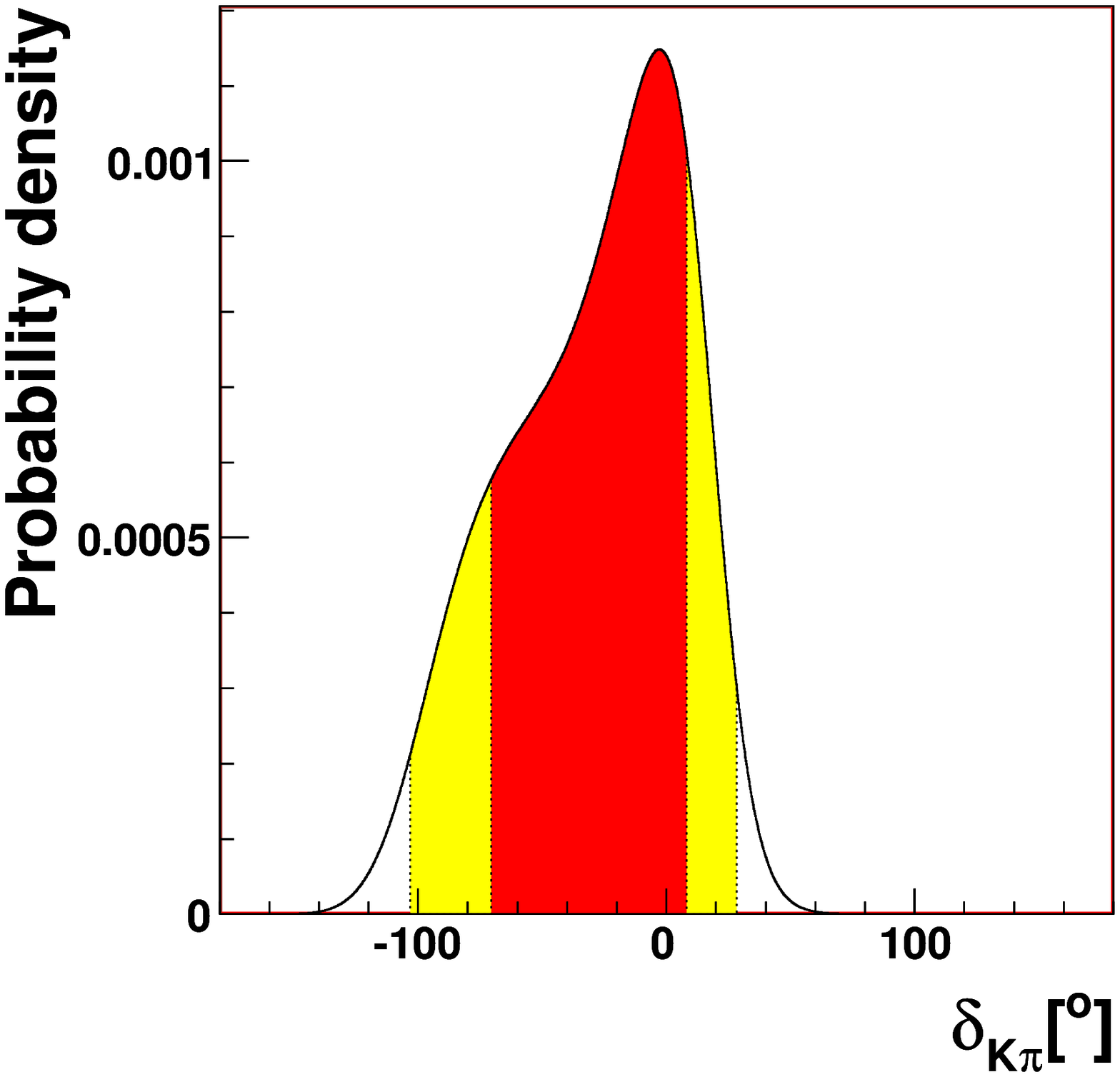}
\includegraphics[width=0.23\textwidth]{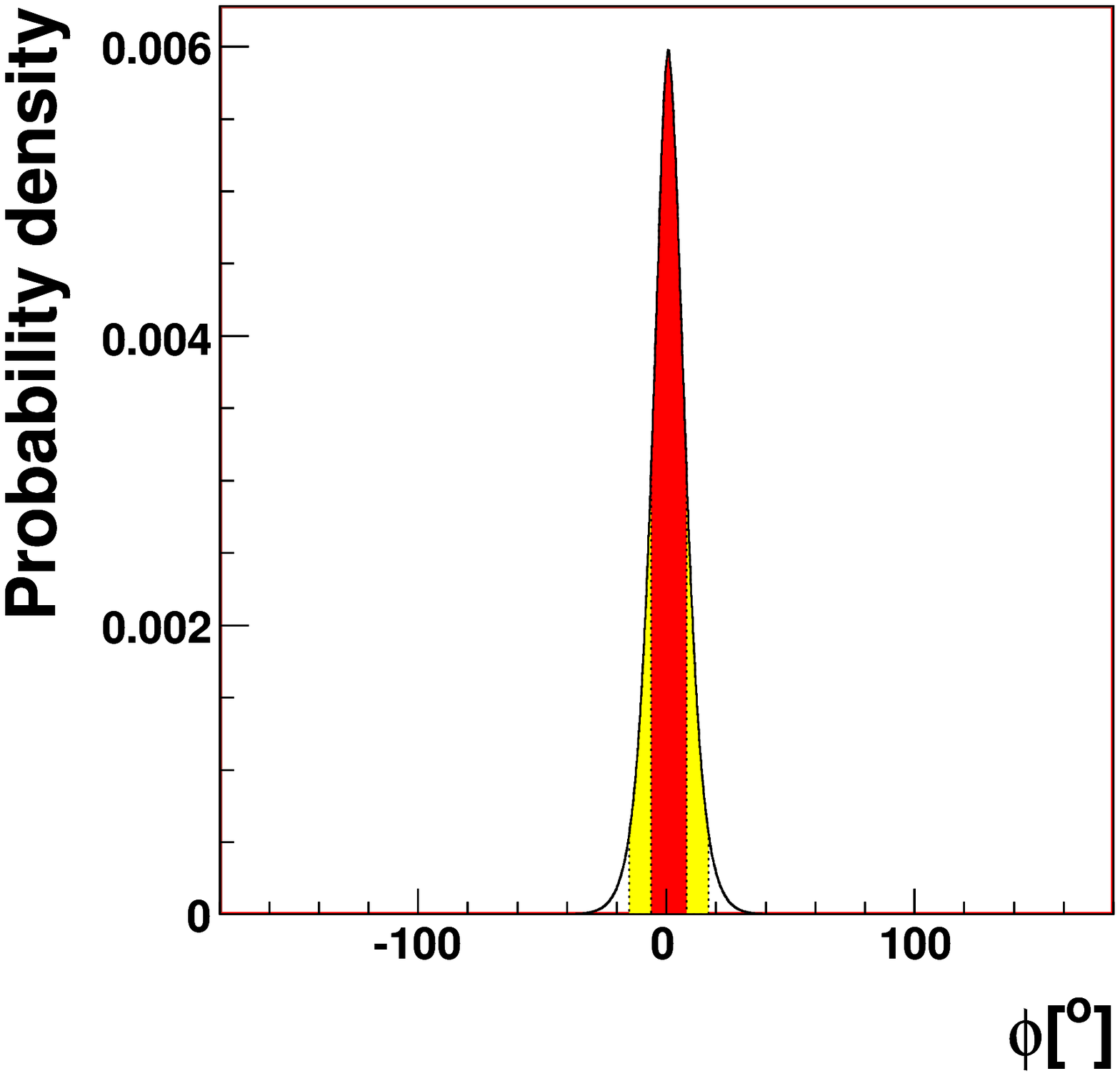}
\includegraphics[width=0.23\textwidth]{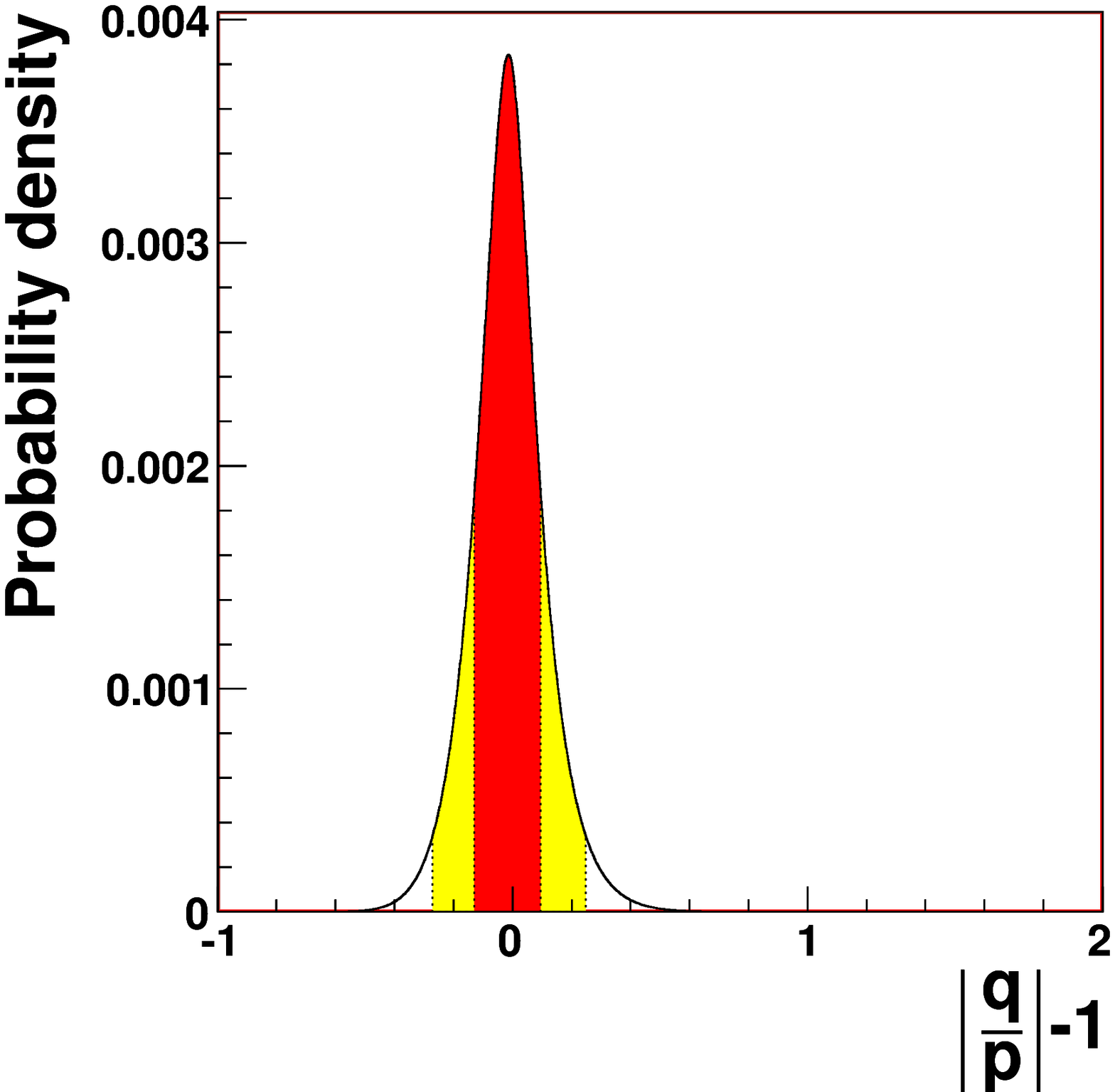}
\includegraphics[width=0.23\textwidth]{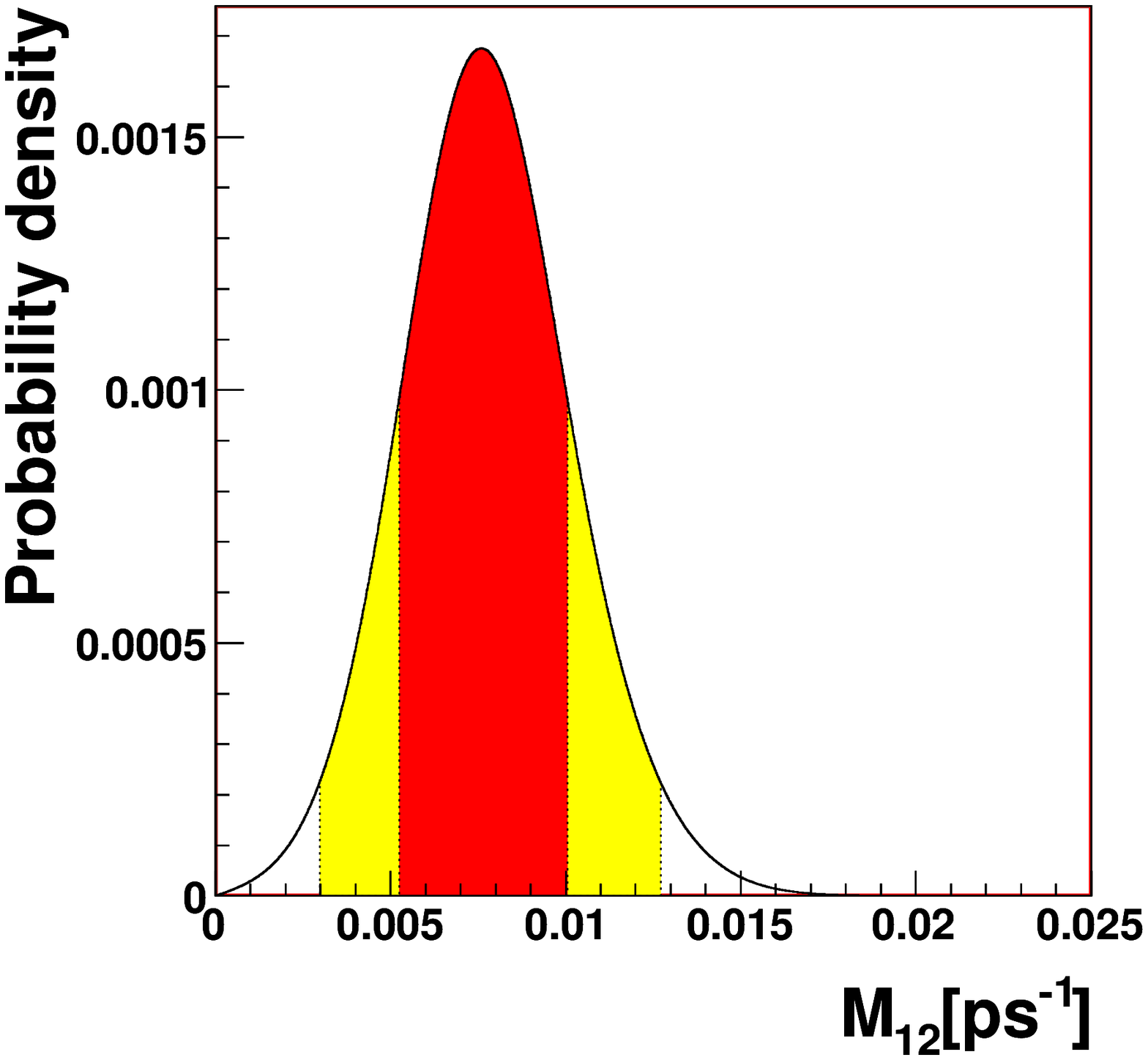}
\includegraphics[width=0.23\textwidth]{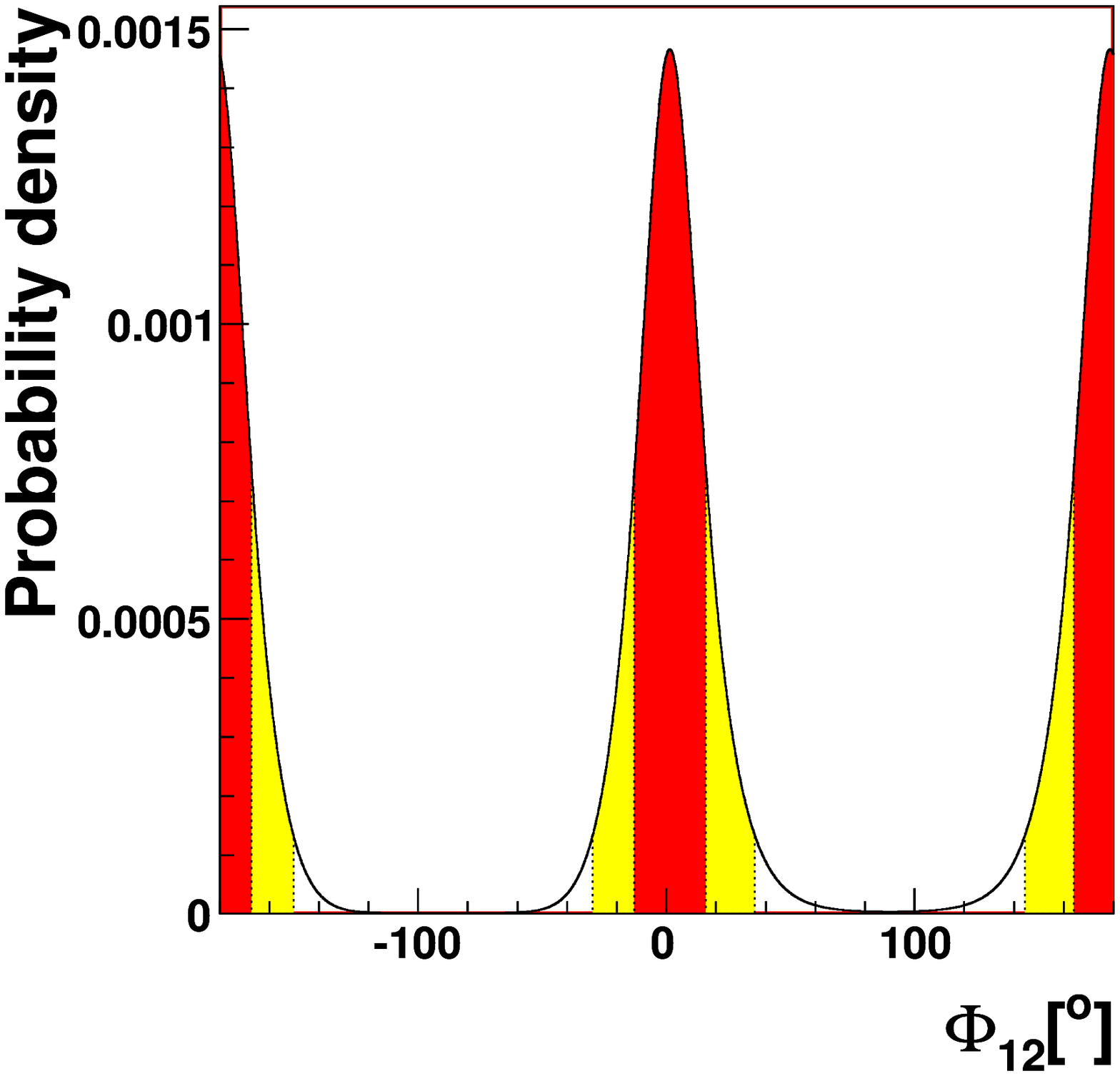}
\includegraphics[width=0.23\textwidth]{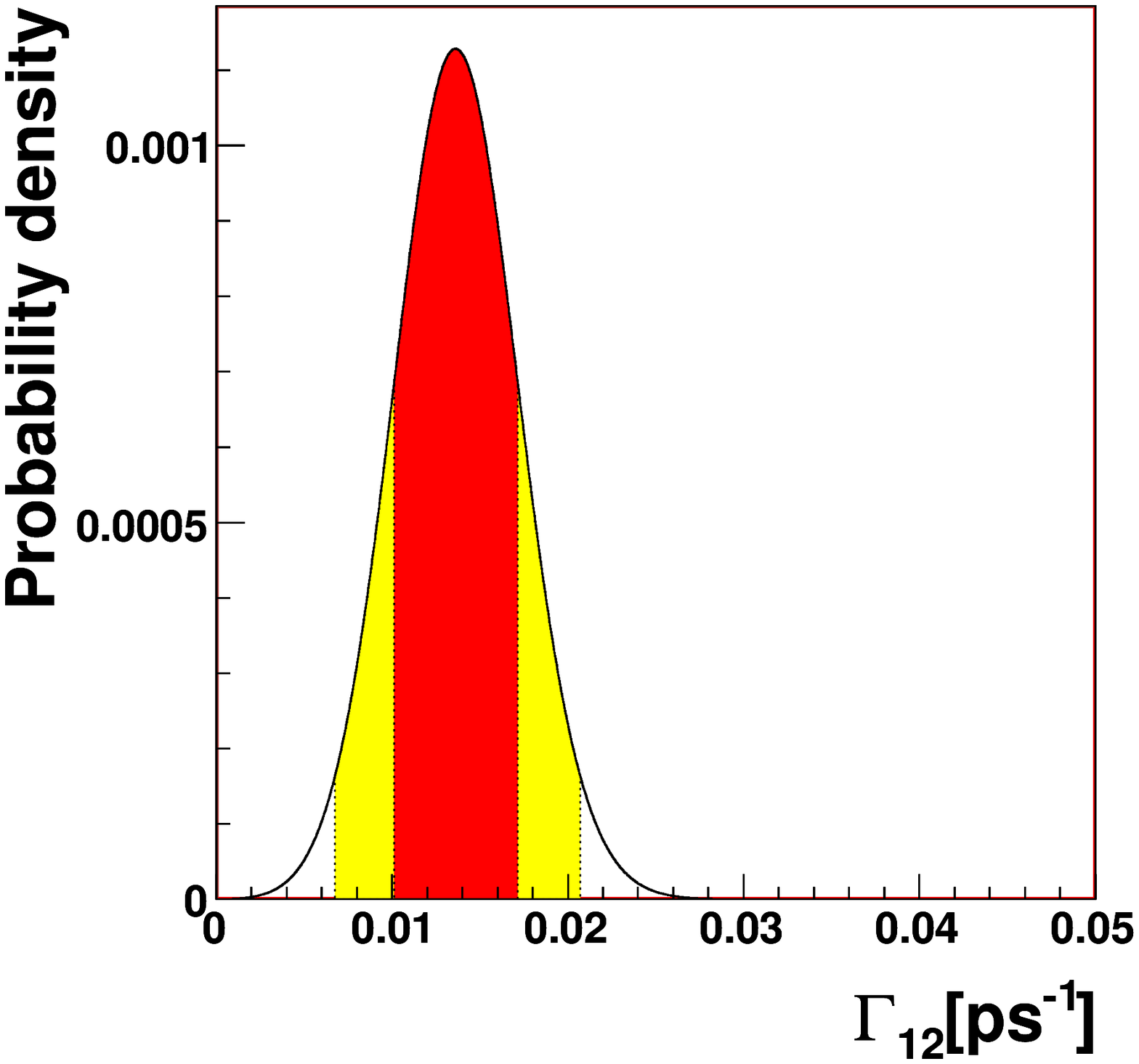}
\caption{%
  Probability density functions of the combined fit from
  Tab.~\protect\ref{tab:exp}. Dark (light) regions correspond to
  $68\%$ ($95\%$) probability.}
\label{fig:1D}
\end{center}
\end{figure}

\begin{figure}[htb!]
\begin{center}
\includegraphics[width=0.22\textwidth]{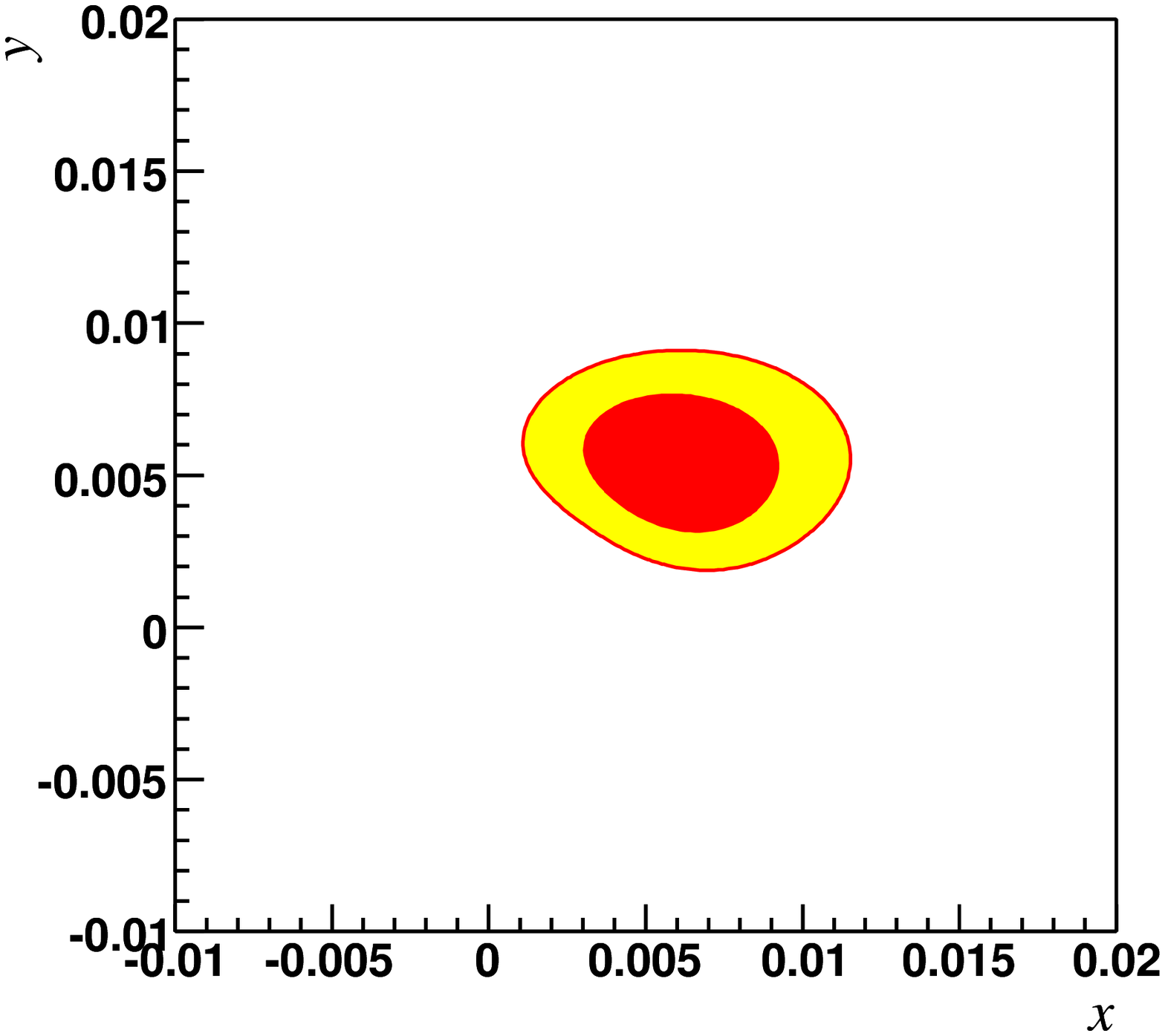}
\includegraphics[width=0.22\textwidth]{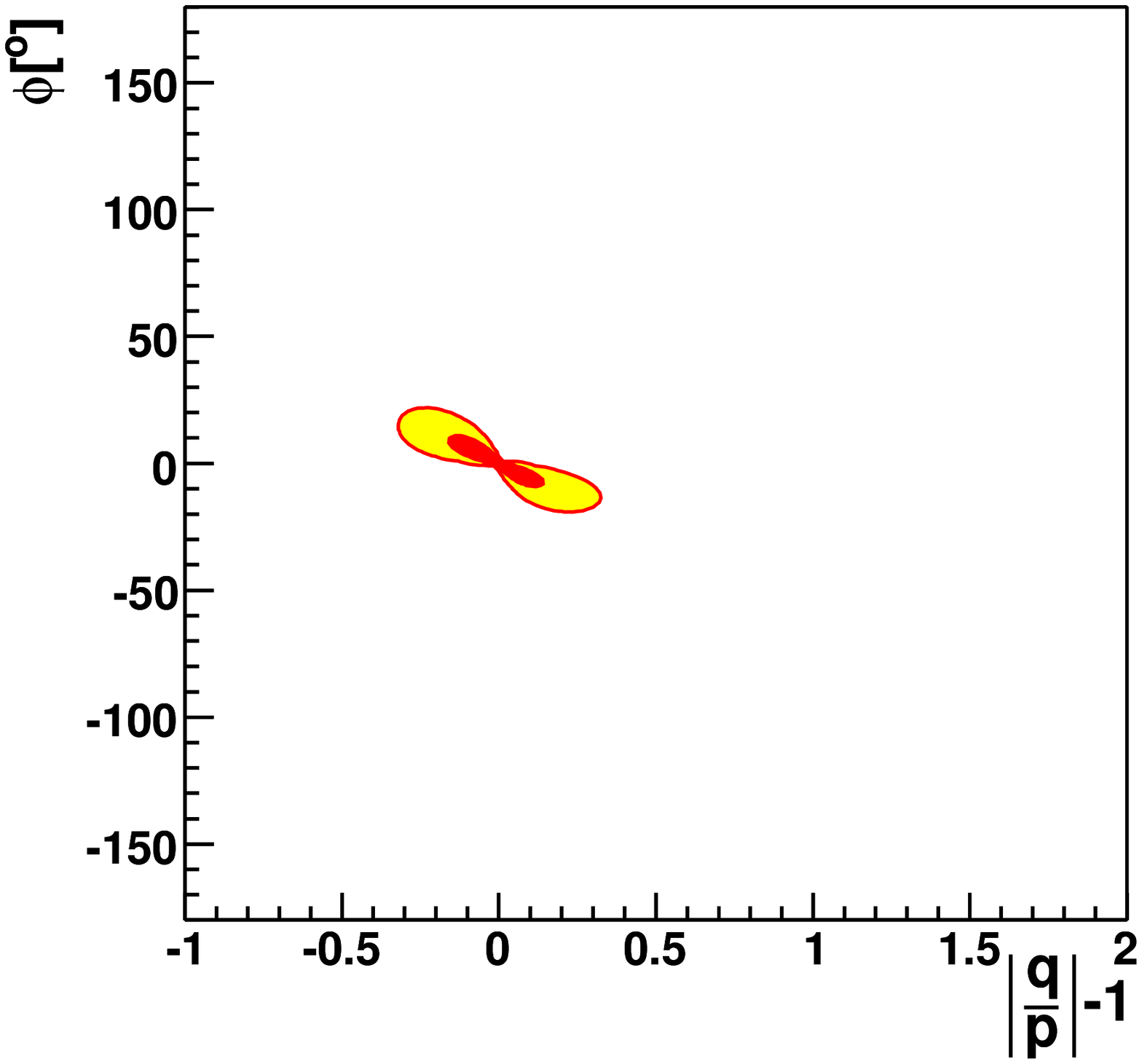}
\includegraphics[width=0.22\textwidth]{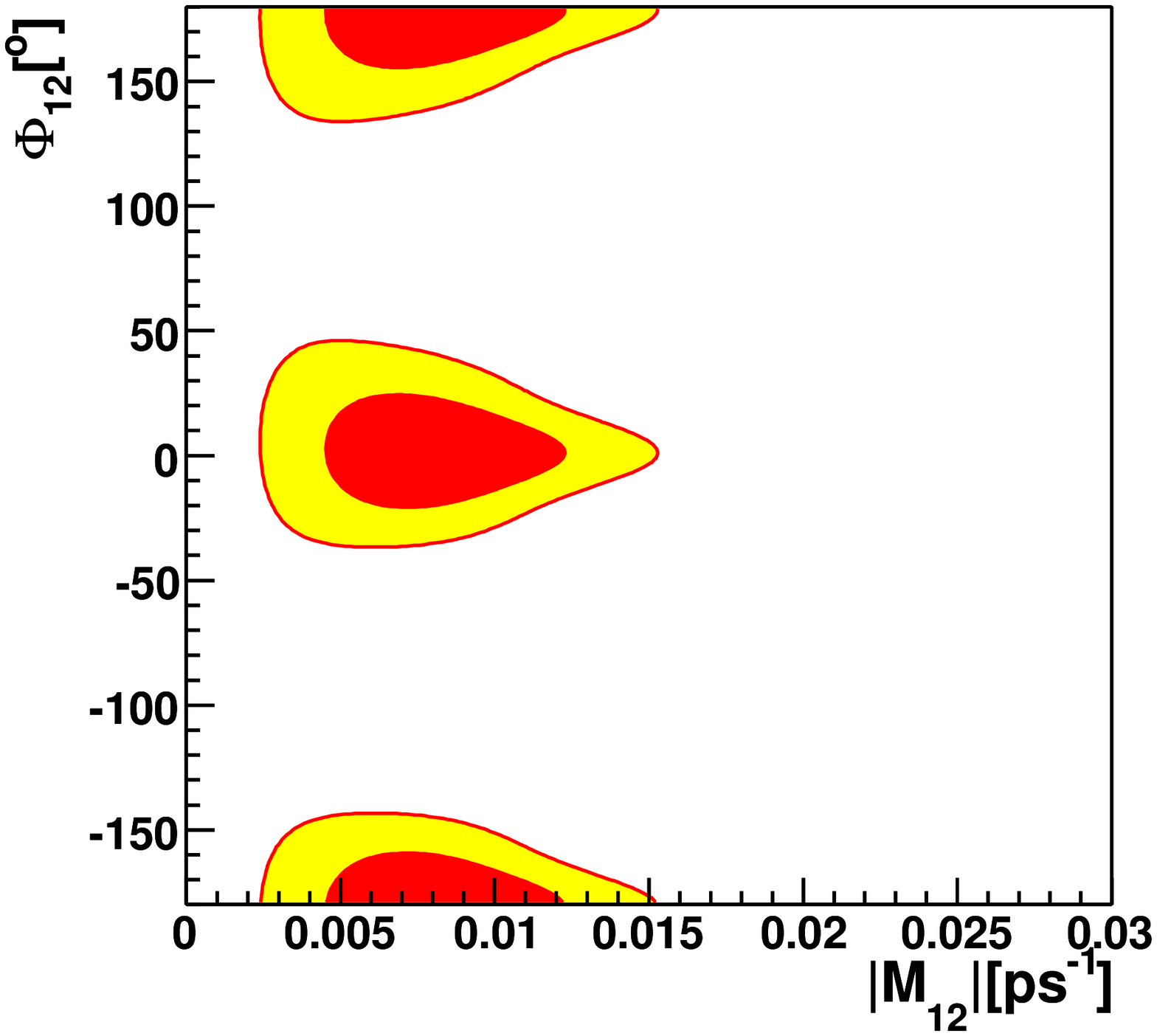}
\includegraphics[width=0.22\textwidth]{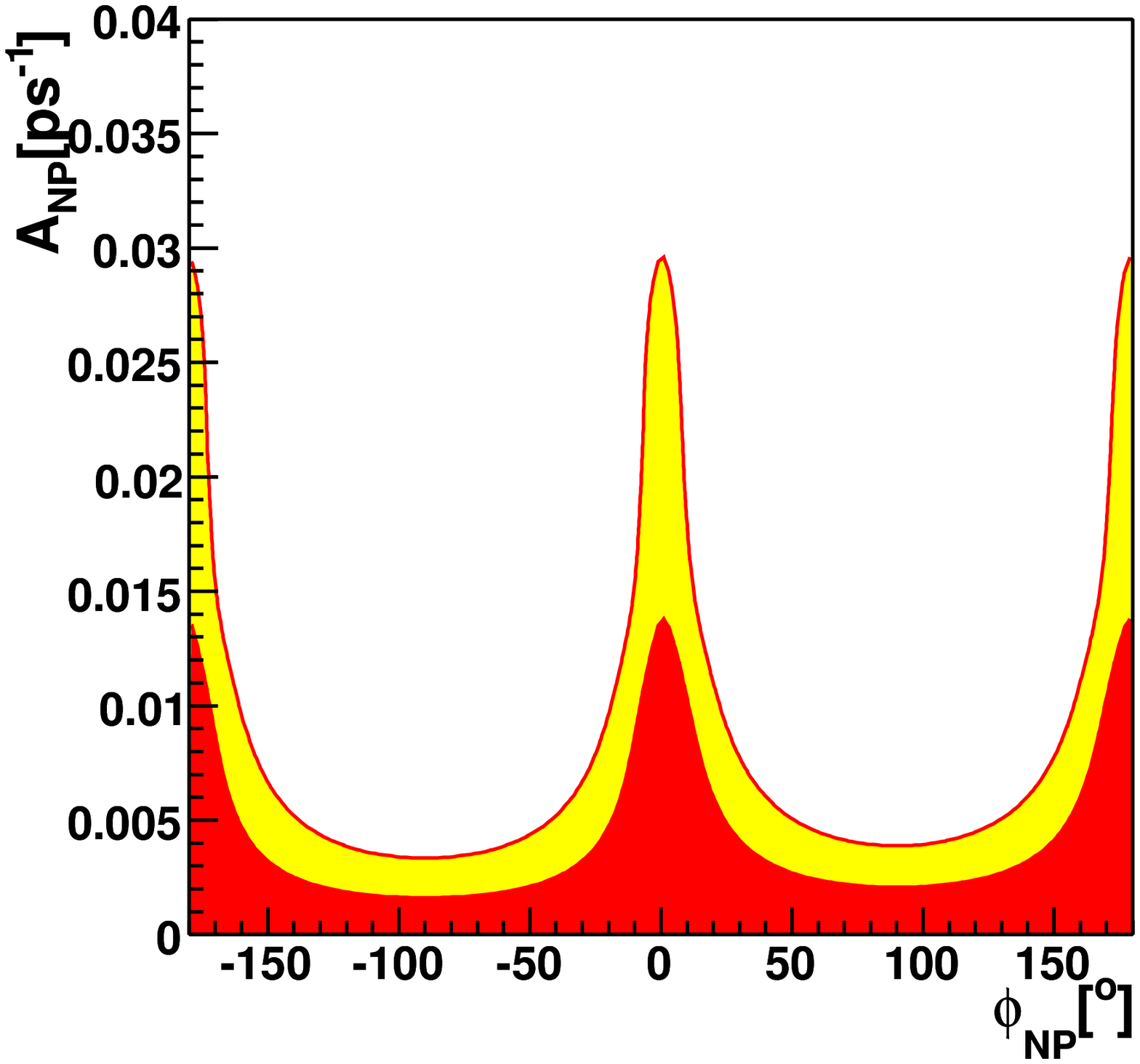}
\includegraphics[width=0.22\textwidth]{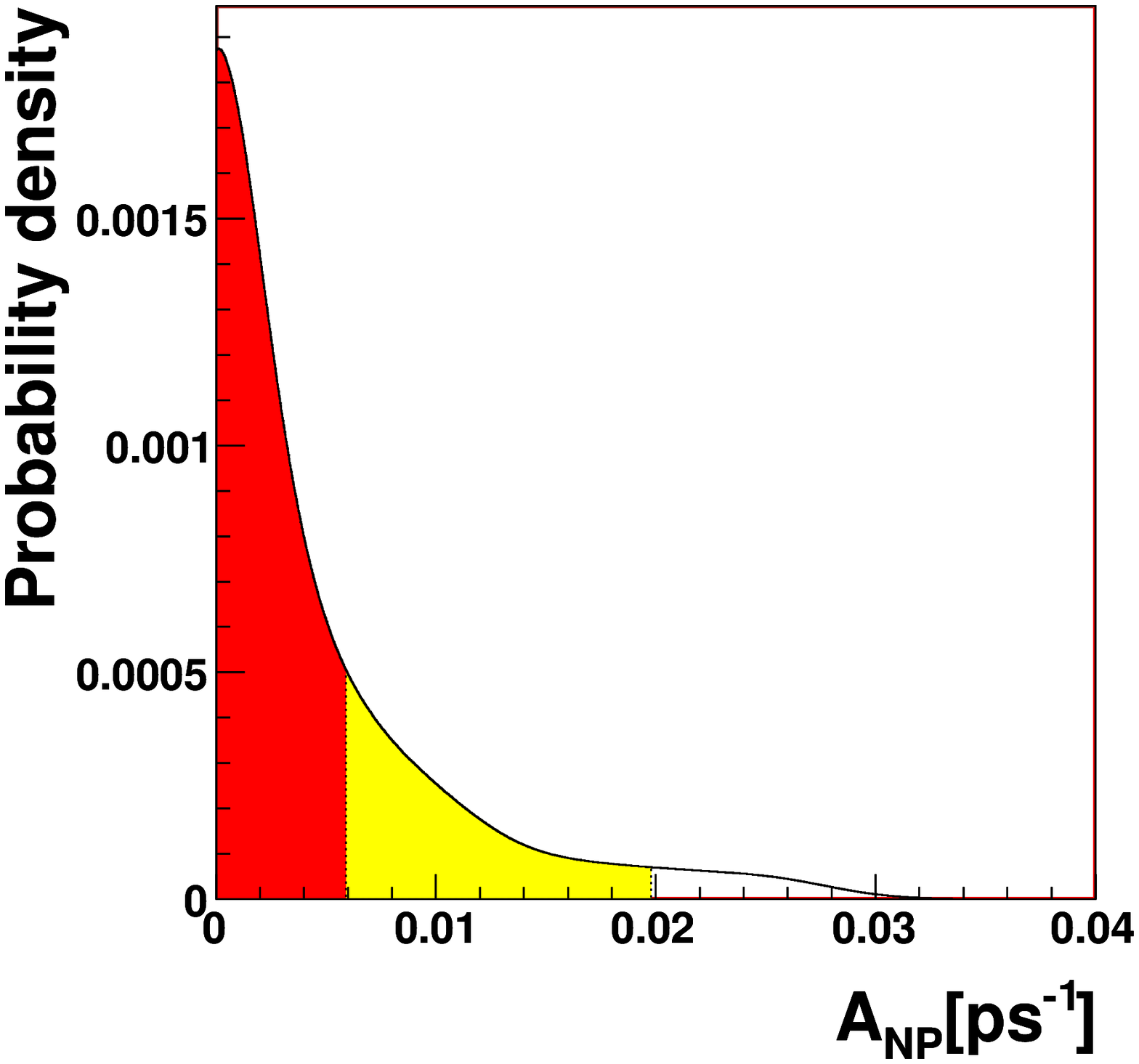}
\includegraphics[width=0.22\textwidth]{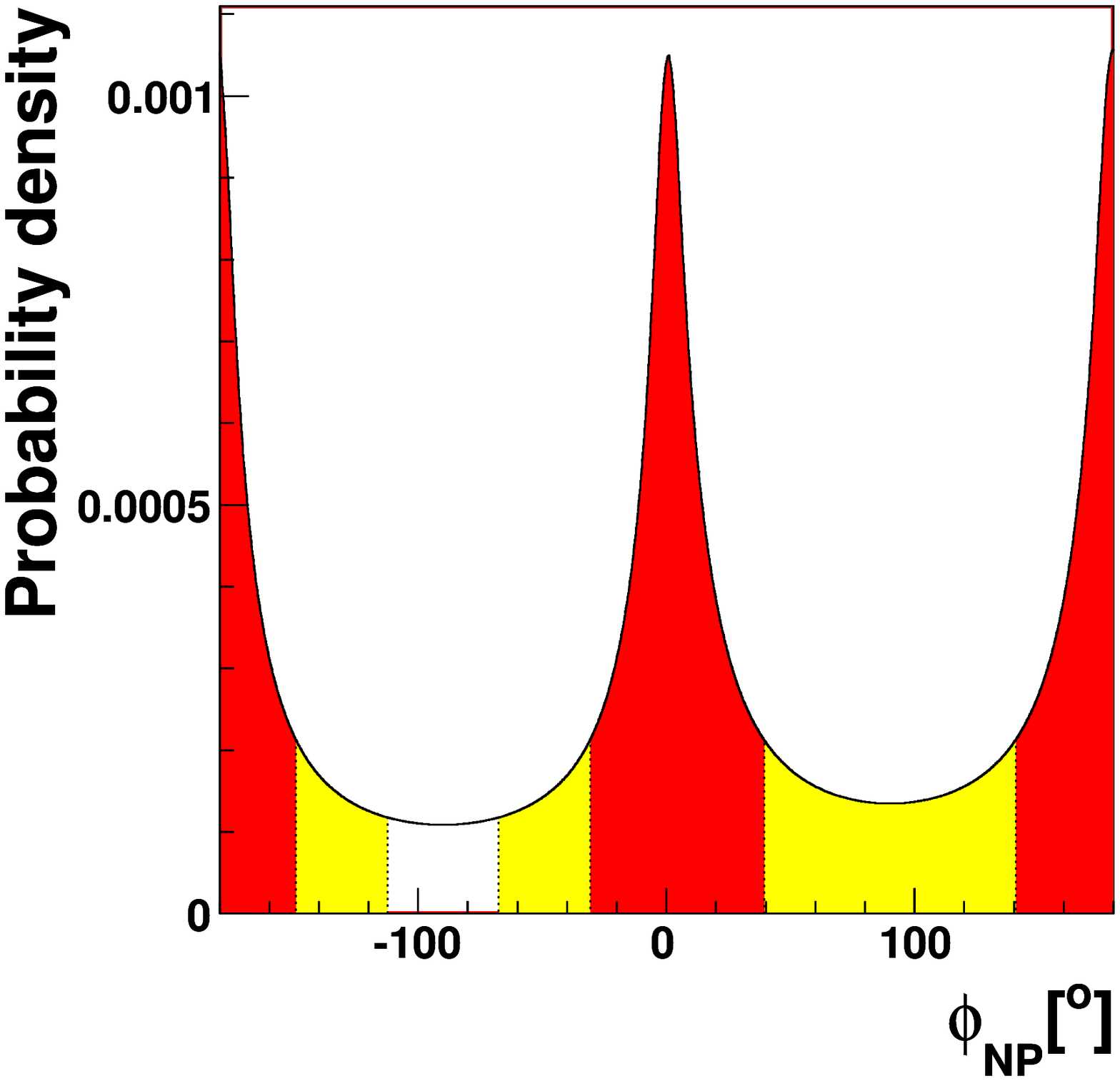}
\caption{%
  Probability density functions of the combined fit from
  Tab.~\protect\ref{tab:exp}, projected on $y$ vs $x$ (top left),
  $\phi$ vs $\vert q/p \vert-1$ (top right), $\Phi_{12}$ vs $\vert
  M_{12} \vert$ (center left), $A_\mathrm{NP}$ vs $\phi_\mathrm{NP}$
  (center right), $A_\mathrm{NP}$ (bottom left) and $\phi_\mathrm{NP}$ (bottom right).
  Dark (light) regions correspond to $68\%$ ($95\%$)
  probability.}
\label{fig:combres}
\end{center}
\end{figure}

The results of the simultaneous fit are quoted in
Tab.~\ref{tab:combres} and shown in Fig.~\ref{fig:1D}. Our
results are slightly different from the HFAG
May 2007 averages because they are obtained allowing for CP violation,
while the HFAG results assume no CP violation. In
Fig.~\ref{fig:combres}, the two-dimensional constraints on the $y$ vs
$x$, $\phi$ vs $\vert q/p \vert$ and $\Phi_{12}$ vs $\vert M_{12}
\vert$ are given.  We notice that, since the measured value of
$y_\mathrm{CP}$ is large, the phase $\phi$ is constrained to be close
to zero. However, due to the large value of $\Gamma_{12}$, the
constraint on $\Phi_{12}$ is less stringent. Some of the results
collected in Tab.~\ref{tab:combres} can be compared with the existing
literature. Concerning the upper bound on $\vert M_{12} \vert$, we
find an improvement of almost an order of magnitude with respect to
the analysis of ref.~\cite{Raz}, while for $x$ the improvement with
respect to ref.~\cite{Golowich} is about a factor of three.

The calculation of $\vert M_{12} \vert$ is plagued by long-distance
contributions~\cite{petrov}.  To take them into account, we proceed in
the following way. We assume that the full amplitude $M_{12}$ is the
sum of the NP amplitude $A_\mathrm{NP} e^{i \phi_\mathrm{NP}}$ and of
a SM real amplitude containing both short- and long-distance
contributions, $A_\mathrm{SM}$. We take $A_\mathrm{SM}$ to be flatly
distributed in the range $[-0.02,0.02]$ ps$^{-1}$, so that it can
saturate the experimental bound in Tab.~\ref{tab:combres}, and derive
from the $\Phi_{12}$ vs
$\vert M_{12}\vert $ distribution the p.d.f. for $A_\mathrm{NP}$ vs
$\phi_\mathrm{NP}$, barring accidental order-of-magnitude cancellations
between SM and NP
contributions. The results, reported in Tab.~\ref{tab:npres} and shown in Fig.~\ref{fig:combres}, provide a
new constraint that should be fulfilled by any extension of the SM.
We see that the lack of knowledge of the SM contribution causes a
dilution of the bound on $\phi_\mathrm{NP}$.  Clearly, if a reliable
estimate of $A_\mathrm{SM}$ were available, the constraint would be
much more effective. Notice also that if $\vert M_{12}\vert$ is
dominated by NP, then $\phi_\mathrm{NP}\sim\Phi_{12}$ and the NP phase
can be experimentally accessed.

\begin{table}[!tb]
\begin{center}
\begin{tabular}{ccc}
\hline
Parameter & 68\% prob. & 95\% prob. \\
\hline
$x$                            & $(6.2  \pm 2.0)\cdot 10^{-3}$  &
$[0.0022,0.0105]$\\ 
$y$                            & $(5.5  \pm 1.4)\cdot 10^{-3}$  &
$[0.0027,0.0084]$\\ 
$\delta_{K\pi}$                & $(-31  \pm 39)^\circ$          &
$[-103^\circ,28^\circ]$\\ 
$\phi$                         & $(1   \pm 7)^\circ$          &
$[-15^\circ,17^\circ]$\\ 
$|\frac{q}{p}|-1$            & $-0.02 \pm 0.11$                &
$[-0.27,0.25]$\\
$\vert M_{12}\vert$ (ps$^{-1}$)  & $(7.7  \pm 2.4)\cdot 10^{-3}$
 & $[0.0030,0.0127]$ \\
$\Phi_{12}$ ($^\circ$)           & $(2   \pm 14) \cup (179 \pm 14)$ &
$[-30,36]\cup[144,210]$\\  
$\vert \Gamma_{12} \vert$  (ps$^{-1}$) & $(13.6 \pm 3.5)\cdot 10^{-3}$
 & $[0.0068,0.0207]$\\
\hline
\end{tabular}
\end{center}
\caption{Results on mixing and CP violation parameters.}
\label{tab:combres}
\end{table}

\begin{table}[!tb]
\begin{center}
\begin{tabular}{ccc}
\hline
 & 68\% prob. & 95\% prob. \\
\hline
$A_\mathrm{NP}$ (ps$^{-1}$) & $[0,0.006]$ &  $[0,0.02]$ \\
$\phi_\mathrm{NP}$ ($^\circ$) & $[-180,-149]\,\cup $
& $[-180,-112]\cup [-68,180]$\\
&  $[-31,39]\cup [141,180]$ & \\
\hline
\end{tabular}
\end{center}
\caption{Allowed ranges for the NP amplitude.}
\label{tab:npres}
\end{table}

\begin{table}[!tb]
\begin{center}
\begin{tabular}{ccc}
\hline
$B_1=0.87\pm0.03$ & $B_2=0.82\pm0.03$ & $B_3=1.07\pm0.09$ \\
$B_4=1.08\pm0.03$ & $B_5=1.46\pm0.09$ & \\
\hline
\end{tabular}
\end{center}
\caption{B parameters defined as in ref.~\cite{bpar} interpolated at
the physical $D$ meson mass, renormalized at the scale $\mu=2.8$ GeV
in the Landau-RI scheme.} 
\label{tab:bpar}
\end{table}

\begin{table}[!htb]
\begin{center}
\begin{tabular}{ccccc}
\hline
$m_{\tilde q}$& $m_{\tilde g}$&$
\left\vert
\left(
  \delta_{12}^u \right)_{LL,RR}
\right\vert$ & $
\left\vert
\left(
  \delta_{12}^u \right)_{LR,RL}
\right\vert$ & $
\left\vert
\left(
  \delta_{12}^u \right)_{LL=RR}
\right\vert$ \\
\hline
350 & 350 & 0.033 & 0.0056 & 0.0020 \\
500 & 500 & 0.049 & 0.0080 & 0.0029 \\
1000 & 1000 & 0.10 & 0.016 & 0.0062 \\
500 & 1000 & 0.14 & 0.011 & 0.0044 \\
500 & 350 & 0.032 & 0.0068 & 0.0025 \\
\hline
\end{tabular}
\end{center}
\caption{Upper bounds at $95\%$ probability for $\left\vert
\left(
  \delta_{12}^u \right)_{AB}
\right\vert$ for various values of squark and gluino masses (in GeV).}
\label{tab:SUSY}
\end{table}

We now turn to the MSSM and consider the bounds on $ \left(
  \delta_{12}^u \right)_{AB}$ that can be obtained from the
determination of $A_\mathrm{NP}$ and $\phi_\mathrm{NP}$ discussed
above. To this aim, we focus on gluino exchange and use the full
Next-to-Leading expression for the Wilson coefficients~\cite{SUSYNLO}
and for the renormalization group evolution down to the hadronic scale
of $2.8$ GeV~\cite{NLORGE}. For the matrix elements, we extrapolate the
results of ref.~\cite{bpar} as given in Table~\ref{tab:bpar} (see also
ref.~\cite{Lin:2006vc} for another recent calculation of $B_1$).

\begin{figure}[htb!]
\begin{center}
\includegraphics[width=0.23\textwidth]{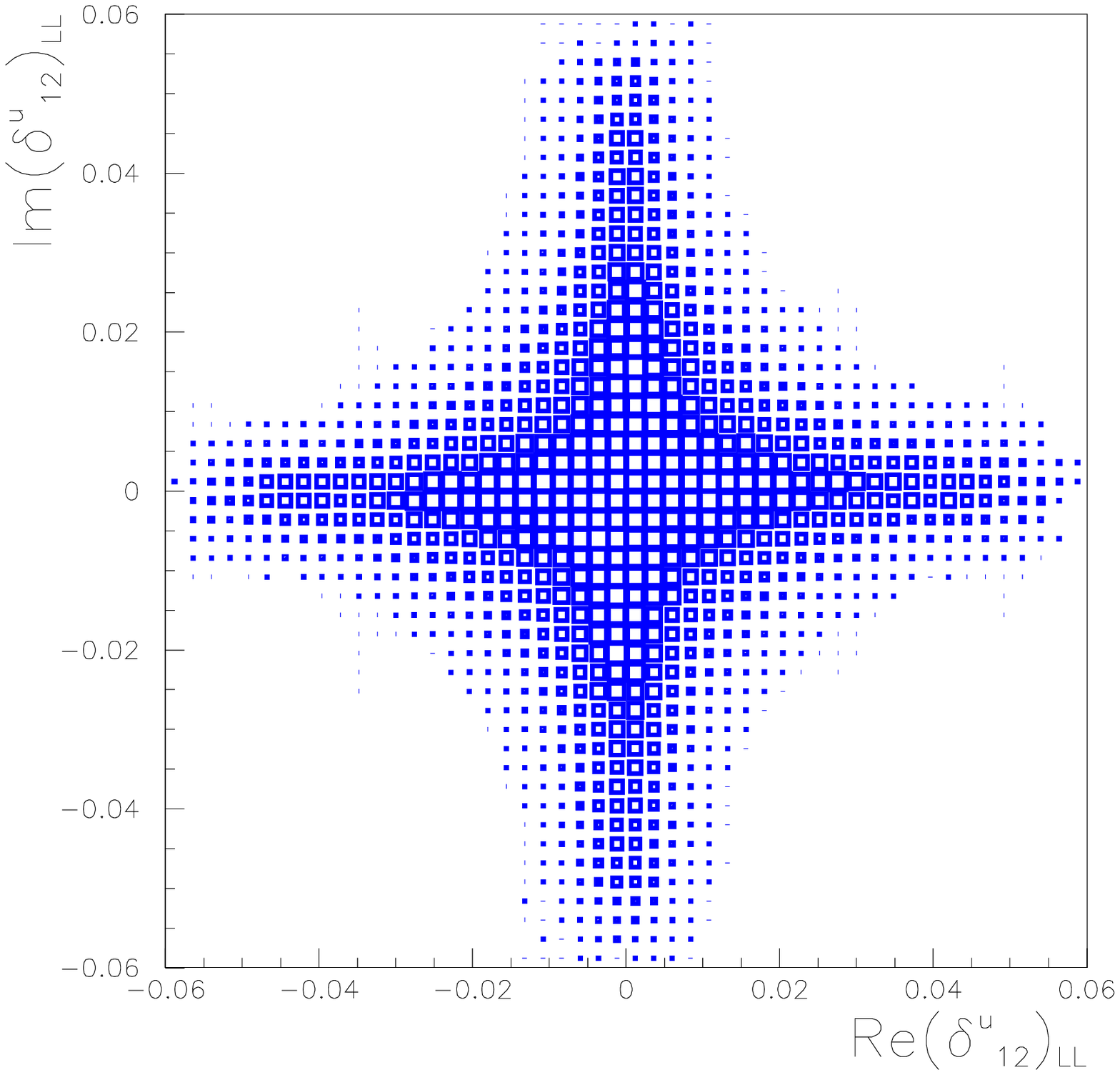}
\includegraphics[width=0.23\textwidth]{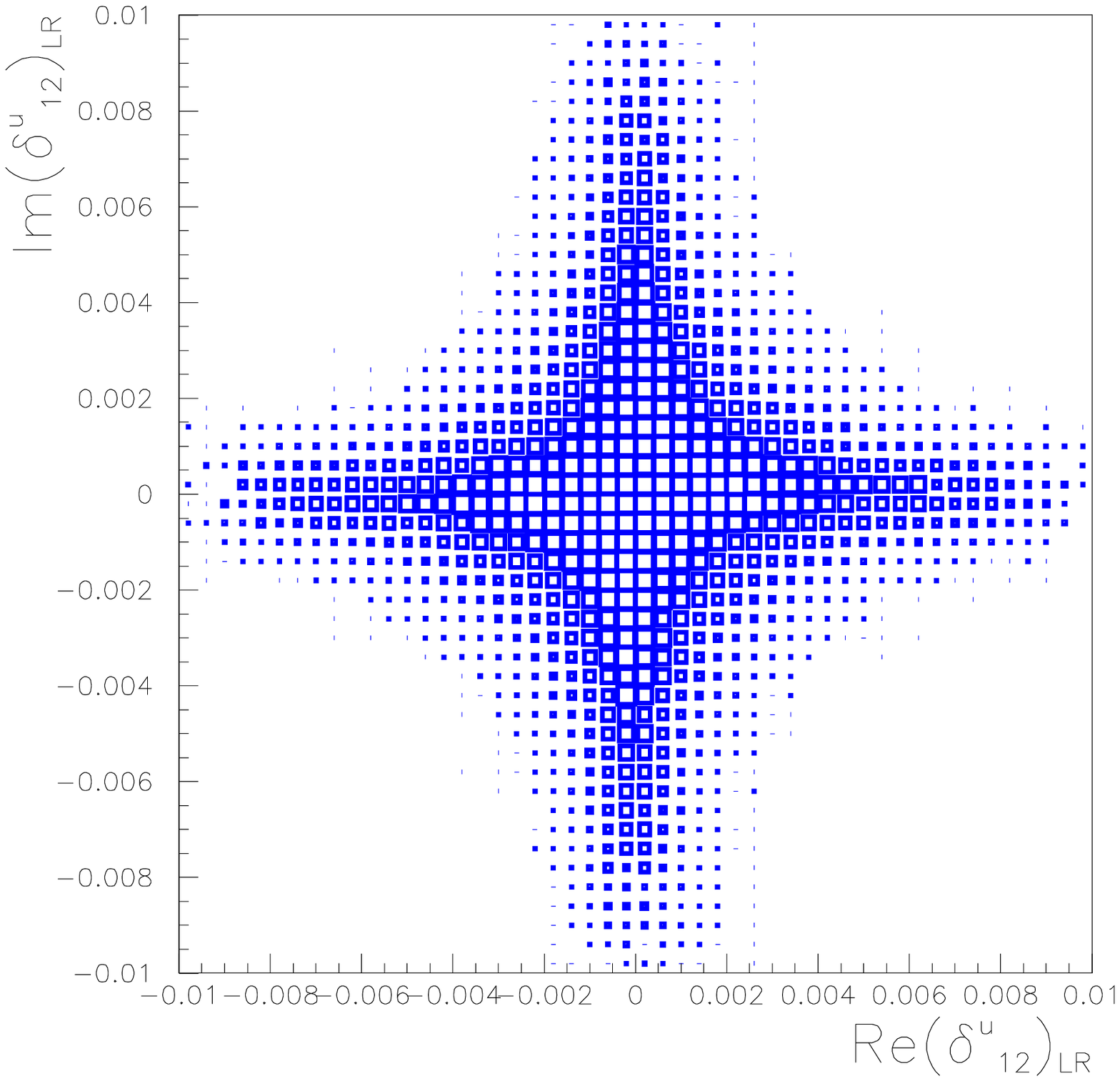}
\includegraphics[width=0.23\textwidth]{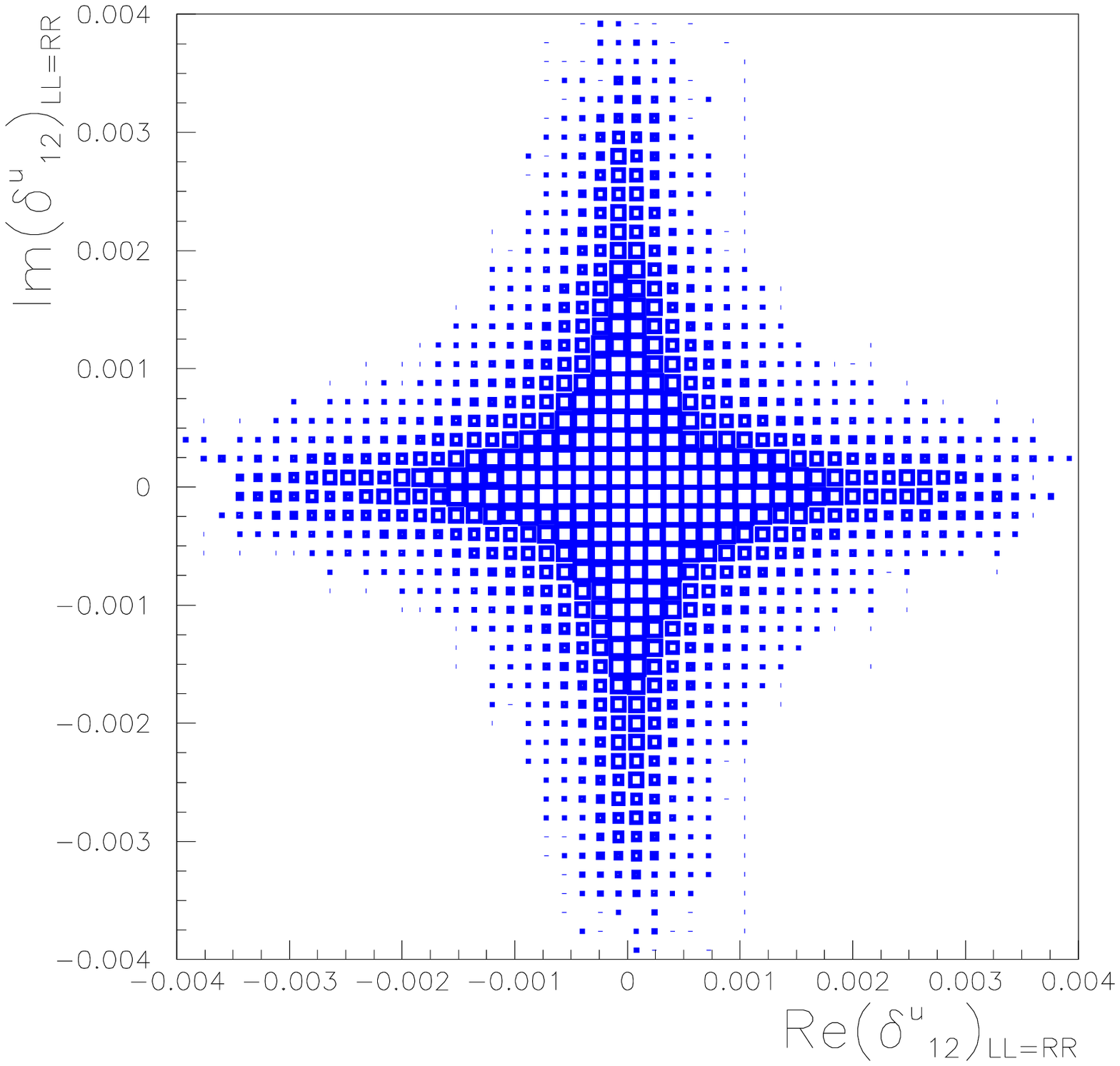}
\includegraphics[width=0.23\textwidth]{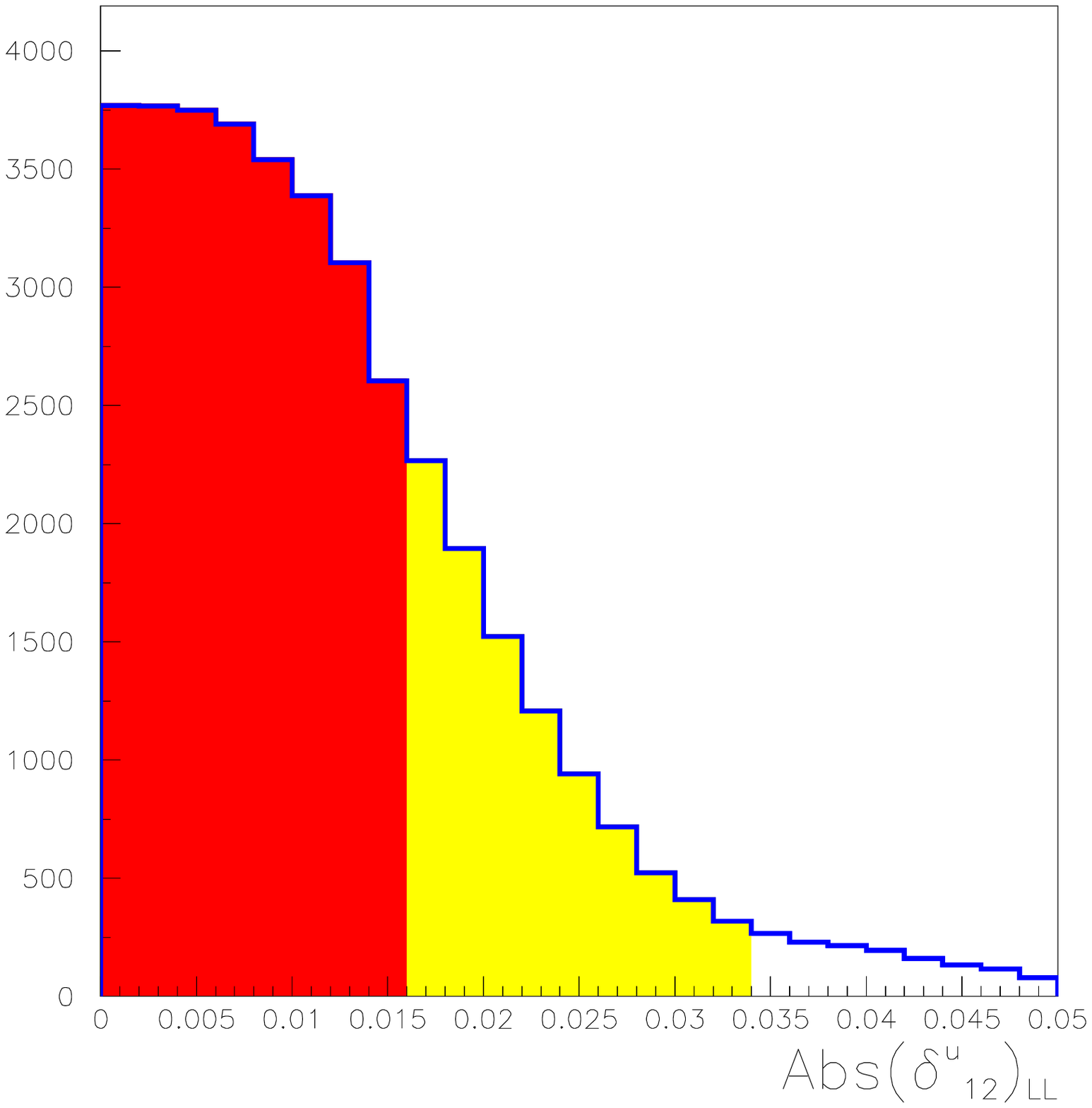}
\includegraphics[width=0.23\textwidth]{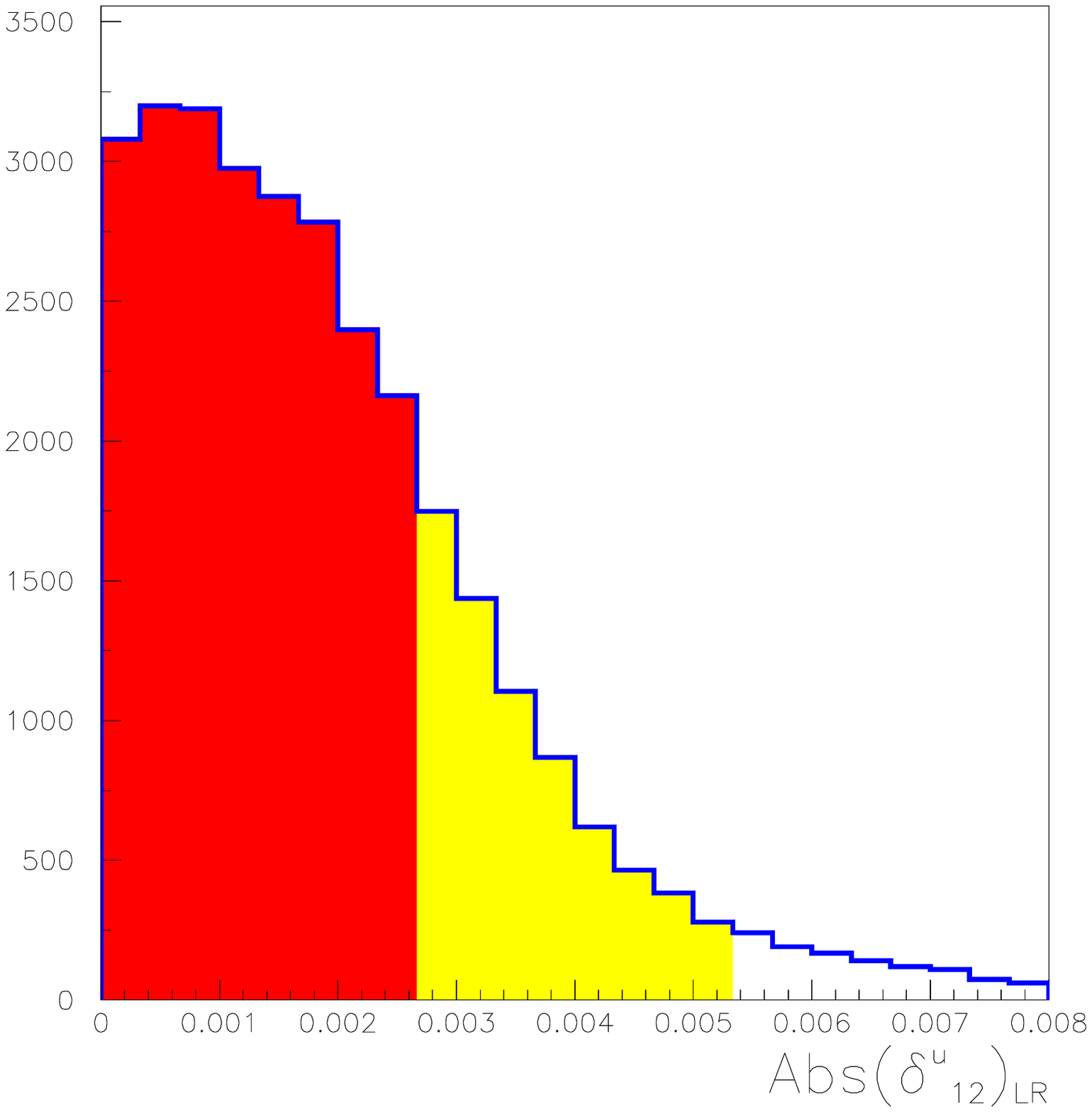}
\includegraphics[width=0.23\textwidth]{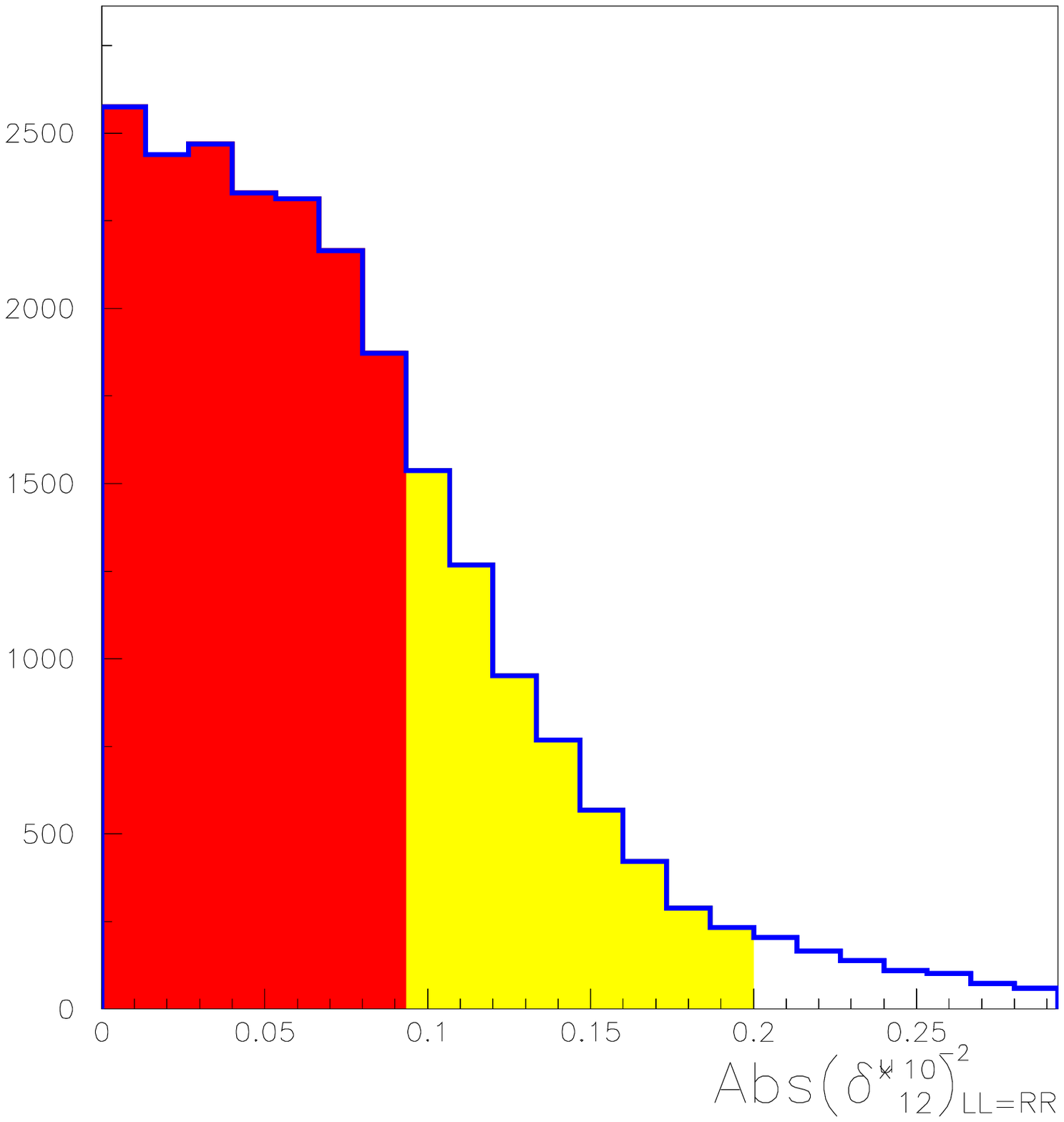}
\caption{%
  Selected regions in the Re$\left(
  \delta_{12}^u \right)_{AB}$--Im$\left(
  \delta_{12}^u \right)_{AB}$ planes and probability density
  functions for Abs$\left(
  \delta_{12}^u \right)_{AB}$ for $AB=LL$, $AB=LR$, $AB=LL=RR$.
  Dark (light) regions correspond to
  $68\%$ ($95\%$) probability. See the text for details.}
\label{fig:SUSY}
\end{center}
\end{figure}

To select the allowed regions on the Re$\left( \delta_{12}^u
\right)_{AB}$--Im$\left( \delta_{12}^u \right)_{AB}$ planes, we use
the method described in ref.~\cite{BsBsbarsusy}. The results are
presented in Fig.~\ref{fig:SUSY} for a reference value of $350$ GeV
for squark and gluino masses. We consider three cases. First, a
dominant $LL$ mass insertion. The case of a dominant $RR$ insertion is
completely identical. Second, a dominant $LR$ insertion. In this case,
chirality-flipping four-fermion operators are generated. These
operators are strongly enhanced by the renormalization group evolution
\cite{bagger}, so that these mass insertions are more strongly
constrained than $LL$ or $RR$ ones. Constraints on $RL$
insertions are identical. Finally, we can switch on simultaneously
$\left( \delta_{12}^u \right)_{LL}=\left( \delta_{12}^u \right)_{RR}$.
In this case, we also generate chirality-flipping operators, so that
the constraint is much stronger than the case in which $\left(
  \delta_{12}^u \right)_{LL} \gg \left( \delta_{12}^u \right)_{RR}$.

In Table \ref{tab:SUSY} we report the bounds on the absolute value of
the mass insertions for several values of gluino and squark masses.
Our bounds are typically a factor of $\sim 3$ more stringent than
those of ref.~\cite{nirraz}.

It is very interesting that SUSY models with quark-squark alignment
generically predict $\left( \delta_{12}^u \right)_{LL}\sim
0.2$~\cite{nirraz}. We conclude that, to be phenomenologically viable,
they need squark and gluino masses to be above $\sim 2$ TeV.
Therefore, they probably lie beyond the reach of the LHC.

In this Letter, we have analyzed the first experimental evidence of
$D$--$\bar D$ mixing recently obtained by the BaBar and Belle
collaborations. Combining the experimental results we obtained new
constraints on the mixing amplitude and on NP contributions. We
have then considered the MSSM and derived new bounds on off-diagonal
squark mass terms connecting up and charm squarks. Finally, we have
briefly commented on the impact of these new constraints on SUSY
models with quark-squark alignment.

We acknowledge partial support from RTN European contracts
MRTN-CT-2004-503369 ``The Quest for Unification'', MRTN-CT-2006-035482
``FLAVIAnet'', MRTN-CT-2006-035505 ``Heptools'' and
fundings from spanish MEC and FEDER under grant FPA2005-01678.

\end{document}